\documentclass[twocolumn,aps,prl,showpacs,amsmath,amssymb,superscriptaddress]{revtex4-1}
\usepackage{graphicx}
\usepackage{dcolumn}
\usepackage{epstopdf}
\usepackage[usenames]{color}

\newcommand{\be}{\begin{equation}}
\newcommand{\ee}{\end{equation}}

\newcommand{\br}{{{\bf{r}}}}

\newcommand{\bq}{{\bf{q}}}

\newcommand{\bea}{\begin{eqnarray}}
\newcommand{\eea}{\end{eqnarray}}
\newcommand{\ra}{\rangle}
\newcommand{\la}{\langle}

\newcommand{\upa}{\uparrow}
\newcommand{\dna}{\downarrow}

\newcommand{\dg}{{\dagger}}
\newcommand{\pdg}{{\phantom\dagger}}

\newcommand{\cS}{{\cal S}}
\newcommand{\SF}{{\cal F}}
\newcommand{\SR}{{\cal R}}

\newcommand{\up}[1]{\ensuremath{^{\textrm{\scriptsize{#1}}}}}

\newcommand{\figref}[1]{Fig.\,\,\ref{#1}}
\newcommand{\BFRO}{{Ba$_2$FeReO$_6$}\,\,}
\newcommand{\CFRO}{{Ca$_2$FeReO$_6$}\,\,}

\begin{document}
\title{Spin-orbital locked magnetic excitations in a half-metallic double perovskite \BFRO}
\author{K. W. Plumb}
\author{A. M. Cook}
\author{J. P. Clancy}
\affiliation{Department of Physics, University of Toronto, Toronto, Ontario, Canada M5S 1A7}
\author{A. I. Kolesnikov}
\affiliation{Neutron Scattering Sciences Division, Oak Ridge National Laboratory, Oak Ridge, TN 37831, USA}
\author{B.~C.~Jeon}
\author{T. W. Noh}
\affiliation{Center for Functional
Interfaces of Correlated Electron Systems, Institute for Basic Science, and
Department of Physics \& Astronomy, Seoul National University, Seoul 151-
747, Korea}
\author{A. Paramekanti}
\affiliation{Department of Physics, University of Toronto, Toronto, Ontario, Canada M5S 1A7}
\affiliation{Canadian Institute for Advanced Research, Toronto, Ontario, M5G 1Z8, Canada}
\author{Young-June Kim}
\affiliation{Department of Physics, University of Toronto, Toronto, Ontario, Canada M5S 1A7}

\date{\today}

\begin{abstract}
We present a powder inelastic neutron scattering study of magnetic excitations in Ba$_2$FeReO$_6$, a member of the double perovskite family of materials which exhibit half-metallic behavior and high Curie
temperatures. We find clear evidence of two well-defined dispersing magnetic
modes in its low temperature ferrimagnetic state. We develop a local moment model, which incorporates
the interaction of Fe spins with spin-orbit locked magnetic moments on Re, and show that this captures
our experimental observations. Our study further opens up double
perovskites as model systems to explore the interplay of strong
correlations and spin-orbit  coupling in 5d transition metal oxides.
\end{abstract}



\maketitle

Strong electronic correlations in the 3d and 4d transition metal oxides (TMOs)
lead to such remarkable phenomena as high temperature superconductivity in the
cuprates \cite{Bednorz1986}, colossal magnetoresistance in the manganites
\cite{Jin1994}, and possible chiral superconductivity in the
ruthenates \cite{Maeno1994}. In 5d-TMOs, the traditional viewpoint suggests that
the larger spread of atomic wavefunctions  leads to a smaller local Hubbard
repulsion and a larger overlap between neighboring atomic orbitals, which
cooperate to suppress strong correlation effects. Indeed, simple oxides like ReO$_3$
are good metals \cite{Phillips1971}. This traditional picture has been challenged
by recent work on iridium-based complex
oxides, which
shows that the large spin-orbit (SO) coupling on Ir can split the
t$_{2g}$ crystal field levels, yielding a reduced bandwidth for effective
$j\!\!=\!\!1/2$ electrons and the re-emergence of strong
correlations \cite{BJKim2008}. Iridates like Na$_2$IrO$_3$, Na$_4$Ir$_3$O$_8$,
Eu$_2$Ir$_2$O$_7$ and Y$_2$Ir$_2$O$_7$, are of great interest since they may
support correlated SO coupled magnetism or topological phases \cite{Yogesh2010,Yogesh2012,Jackeli2009,Shitade2009,Okamoto2007,Lawler2008,Yanagishima2001,Witczak2011,Wan2011}.

\begin{figure}[tb] \includegraphics[scale=1.0]{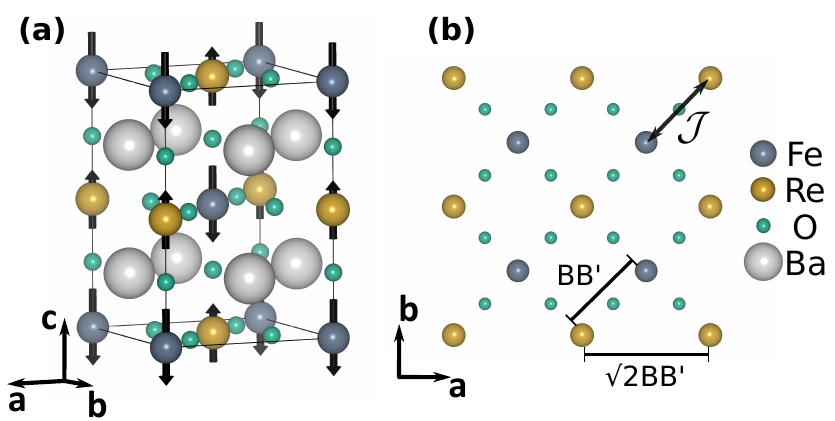}
    \caption{\label{Fig:xtal} Crystal structure of \BFRO\!\!. (a) Schematic of
    the crystallographic unit cell showing the relative orientation of Fe and
    Re moments.  (b) Projection into ab-plane illustrating the separation
    between B(B') site ions, and the exchange pathway $\cal{J}$ between
neighboring Fe and Re sites.} \end{figure}

A distinct route to strong correlations in 5d-TMOs may be realized in {\em
ordered} double perovskites (DPs), with chemical formula A$_2$BB'O$_6$,
obtained by stacking alternating ABO$_3$ and AB'O$_3$ perovskite units. In this
structure, neighboring B(B') site ions are further apart by
$\sqrt{2}$ (see Fig.~\ref{Fig:xtal}). With 5d metal ions on the B'
sites, the larger B'-B' distance suppresses direct 5d orbital overlap,
enhancing strong correlations. There is a growing interest in DPs such as
A$_2$FeReO$_6$ (A=Ca,Sr,Ba)
\cite{Longo1961,Sleight1962,Kobayashi1999,Serrate2007,Azimonte2007,Jeon2010,Garcia2012}
and Sr$_2$CrOsO$_6$ \cite{Krockenberger2007,Erten2012}, with a 3d magnetic B-ion and
a 5d B'-ion, since they exhibit high Curie temperatures. In addition, the half-metallic character
and significant polarization of
many DPs makes them ideal candidates for spintronic applications such as spin injection \cite{Kobayashi1998,Zutic2004}.
Despite this great fundamental
and technological interest in the DP materials \cite{Jackeli2003,Harris2004,Brey2006,Erten2011}, there is a significant lack of experimental
work on their magnetic excitations.

\begin{figure}[tb]
    \includegraphics[scale=1.0]{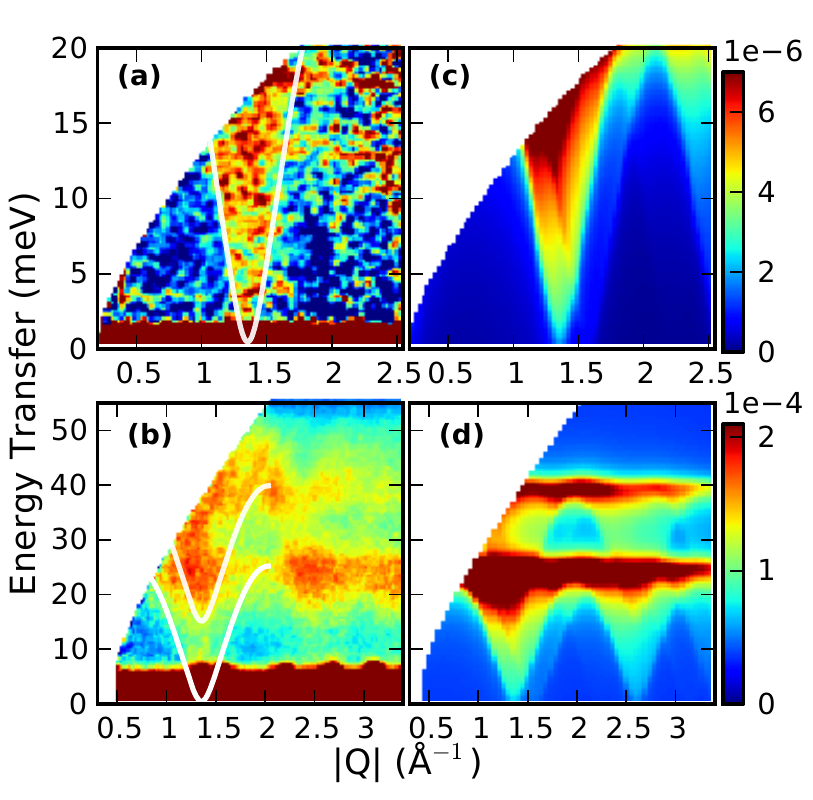}
    \caption{\label{Fig:SQW} Comparison of spin wave model and measured powder
    averaged magnetic scattering intensity for T = 34 K. An empty Al-Can
    background signal has been subtracted from the data. (a)  and (b) are
    neutron scattering data for incident energies of 27 and 120 meV
    respectively.  (c)-(d) Powder averaged dynamic structure factor calculated
   from the spin wave model with $6  {\cal J}_{\rm eff} \SF=39$~meV, and
    $6 {\cal J}_{\rm eff} \SR = 25$~meV. Solid white lines in (a) and (b) show the
    dispersion relation from the spin wave model along $(0,0,0) \text{--}
    (\pi,\pi,\pi) \text{--} (3\pi/2,3\pi/2,3\pi/2)$.}
\end{figure}

In this Letter, we address this issue by using inelastic neutron scattering (INS) on
polycrystalline Ba$_2$FeReO$_6$ to study the magnetic excitations in its
ferrimagnetic state, complemented by a theoretical modelling of the observed
spectrum.  Our main results, which are summarized in Fig.~\ref{Fig:SQW}, are as
follows. (i) We provide experimental evidence of {\em two} dispersive magnetic
modes in the magnetic excitation spectrum, showing that Fe and Re electrons
both exhibit strong correlations and contribute to the magnetization dynamics.
(ii) We find evidence of nearly gapless magnetic excitations in the inelastic
spectrum, indicating a weak locking of Re-moments to the lattice in the
ferrimagnetic state. (iii) We discuss a minimal local moment model of strongly
coupled spin and orbital degrees of freedom on Re interacting with spins on Fe,
which captures our experimental observations.  (iv) We combine our results with
published magnetization and X-ray magnetic circular dichroism (XMCD) data to
obtain estimates of the Re and Fe moments and the effective Re-Fe exchange
interaction.  Our work further opens up 5d-based DPs as model systems to study
the interplay of spin orbit coupling and strong electronic correlations.

\noindent {\bf Experiments:}
Total $8.6$g of polycrystalline \BFRO\ sample was synthesized
using the standard solid-state method reported previously \cite{Jeon2010,Prellier2000}.
In some DPs, anti-site disorder (mixing of B
and B' site atoms) is significant, and suppresses saturated magnetic moments.
However, for \BFRO, a large difference ($\sim$8\%) in the ionic radii of
Fe$^{3+}$ and Re$^{5+}$ seems to mitigate this problem. From the structural refinement
of x-ray powder diffraction data, we infer an anti-site disorder of $\lesssim \! 1$\%,
consistent with that reported by Winkler et al. \cite{Winkler2009}.

Neutron scattering measurements were carried out on the fine resolution
Fermi-chopper spectrometer SEQUOIA at the Spallation Neutron Source (SNS) at
Oak Ridge National Laboratory (ORNL). Measurements were performed with Fermi
chopper 1 rotating at a frequency of 300 Hz and phased for incident energies of
27 and 120 meV. A T0 chopper rotating at 180 Hz was used to eliminate a fast
neutron background. The sample was sealed in an Al can and mounted
on a closed cycle cryostat.
Data were also collected for an empty Al sample can at T = $34$~K, with an
identical instrumental configuration. The absorption corrected empty can intensities were subtracted from the raw
data at T = $34$~K to remove scattering from the sample environment.

Throughout this article we use pseudo-cubic notation $a=b=c \approx 4.01$~\AA\,
and index the momentum transfer $\mathbf{Q}$ in units of $1/a$ to aid
comparison with theoretical calculations. In our magnetic model ferrimagnetism
arises from G-type antiferromagnetic arrangement of inequivalent Fe and Re
moments so that magnetic Bragg peaks occur at the antiferromagnetic wavevector
$\mathbf{Q}_{\mathrm{AF}} \!=\!  (\pi,\pi,\pi)$ and the ferromagnetic
wavevector $\mathbf{Q}_{\mathrm{FM}} \!=\!  (2\pi,0,0)$.

Maps of the inelastic neutron scattering intensity for 27~meV and 120~meV
incident neutron energies are shown in \figref{Fig:SQW} (a) and (b)
respectively. An inelastic feature emanating from $Q \!=\!  1.35$~\AA\up{-1}
corresponding to $\mathbf{Q}_{\mathrm{AF}}$ is clearly resolved. The inelastic
feature extends into two-bands of excitations with maximum intensities near
$25$~meV and $39$~meV.  The scattering is strongest at low-Q and decays rapidly
for increasing Q as is expected generally from the form factor dependence for
magnetic scattering.  Results from our theoretical model are shown in
\figref{Fig:SQW} (c)-(d) with the best-fit parameters.

\begin{figure}[tb]
    \includegraphics[scale=1.0]{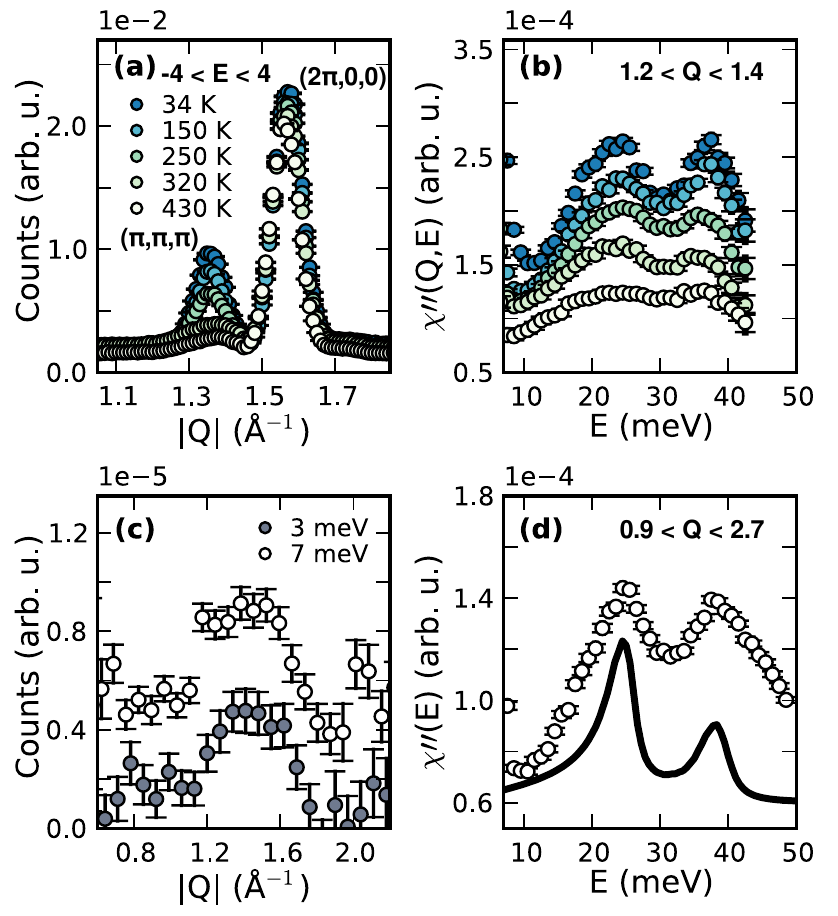}
    \caption{\label{Fig:cuts} Representative cuts through S(Q,E). (a)
Temperature dependence of magnetic Bragg peaks at $\{\pi,\pi,\pi\}$
and  $\{2\pi, 0, 0\}$ determined by integrating the $E_i = 120$~meV data over the
elastic line resolution [$-4 < \text{E} < 4$~meV]. (b) Temperature dependence
of Bose factor corrected inelastic scattering near the magnetic zone center.
(c) Constant energy cuts from the $E_i = 27$~meV data at 34~K, each energy cut
was integrated over $\pm1$ meV, the 7 meV cut is offset for clarity.  (d) Bose
factor corrected inelastic scattering at 34~K integrated over a magnetic
Brillouin zone representing the magnetic density of states, the solid line is
an equivalent cut from the spin wave model calculation. An empty can background
has been subtracted from data in (c) and (d).}
\end{figure}

The temperature and energy dependence of putative magnetic scattering in
\BFRO\, is presented in \figref{Fig:cuts}. Bragg peaks at $Q \! = \!
1.35$~\AA\up{-1} corresponding to $\mathbf{Q}_{\mathrm{AF}}$, and $Q \!=\! 1.56$~\AA\up{-1} corresponding to $\mathbf{Q}_{\mathrm{FM}}$ are shown in
\figref{Fig:cuts} (a). The elastic magnetic intensity decreases upon warming
and the antiferromagnetic Bragg peak vanishes above $300$~K, consistent with
the reported $T_c \approx 304$~K for \BFRO\!\! \cite{Azimonte2007}. Constant
momentum transfer cuts detailing the inelastic scattering emerging from the
magnetic zone center are shown in \figref{Fig:cuts} (b). The
fluctuation-dissipation theorem $S(Q,E) = (n(E,T) + 1)
{\chi}^{\prime\prime}(Q,E)$ relates the imaginary component of the dynamic
susceptibility ${\chi}^{\prime\prime}(Q,E)$ to the dynamic structure factor
measured directly by neutron scattering where $n(E,T)$ is the Bose thermal
occupation factor. Correcting the INS intensity by the Bose factor allows for
comparison of the inelastic scattering across the entire $400$~K temperature
range on a single intensity scale. Two strong inelastic features are visible
near $25$~meV and $39$~meV which decrease in intensity upon increasing
temperature. The temperature, momentum, and energy dependence of the low-Q
inelastic scattering is entirely consistent with expectations for scattering
from powder averaged spin-waves. Broader examination of the data reveals two
bands of phonon scattering which partially obscures the magnetic signal above
$3$~\AA\up{-1}; however, the phonon and magnetic scattering are well resolved
since the magnetic form factor rapidly attenuates the magnetic intensity with
increasing $Q$ while the phonon scattering intensity increases with $Q$ (see
Supplemental Material).  Constant energy cuts across the low energy
magnetic scattering are shown in \figref{Fig:cuts} (c). An inelastic feature
emerging from  the antiferromagnetic zone center is clearly resolved within our
experimental resolution down to at least $3$~meV. The scattering intensity is
strongest near the antiferromagnetic wavevector at $Q\!= \! 1.35$~\AA\up{-1}
--- where the structure factor for magnetic scattering is maximized --- and is
small near the nuclear Bragg peak. This Q-dependence identifies the low energy
inelastic scattering as magnetic in origin and places an upper bound of $3$~meV
for any gap in the spin wave dispersion. The Q-integrated inelastic intensity
is peaked at the magnon zone boundary energy where the density of states for
spin waves is maximized, enabling a precise determination of the zone-boundary
energies from the powder averaged spectrum. The dynamic susceptibility
integrated over the magnetic Brillouin zone is shown in \figref{Fig:cuts} (d);
scattering is strongly peaked at $25$~meV and $39$~meV. An equivalent cut from
the powder averaged spin wave theory using the same parameters as in
\figref{Fig:SQW} (c) and (d) is also shown in the figure.

\noindent {\bf Local moment model:} The well-defined magnetic modes in
Fig.~\ref{Fig:SQW}, and the fact that the closely related material \CFRO
is an insulator, suggests that strong electronic correlations are important in
\BFRO\!\!. A local moment model thus provides a useful vantage point to
describe its magnetic excitations. The simplest such Hamiltonian is $H\!\!=\!\!
{\cal J} \sum_{\la \br\br' \ra} \vec{S}_\br \! \cdot \! \vec{\SF}_{\br'} \! -\!
\lambda \sum_{\br\in Re} \vec{L}_\br \! \cdot \! \vec{S}_\br$, with a
nearest neighbor antiferromagnetic exchange interaction between the Fe spin
$\vec\SF$ and the Re spin $\vec S$ induced by intersite tunneling.
In addition, we include SOC between
orbital ($\vec L$) and
spin ($\vec S$) angular momentum on Re. We ignore SOC on Fe. This
model should be broadly applicable to many DPs with an orbitally nondegenerate
magnetic B-site (e.g., Fe), and a magnetic B'-site with active $t_{2g}$
orbitals (e.g., Re).

\begin{figure}[tb]
\includegraphics[scale=1.0]{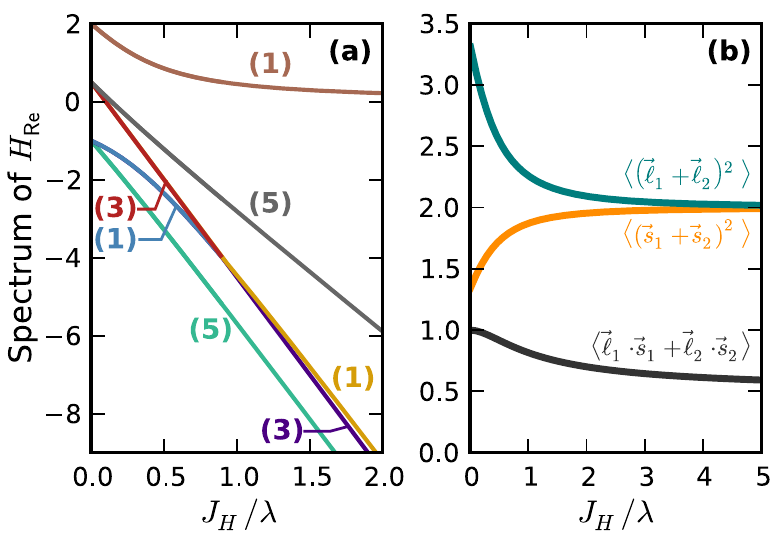}
\caption{(a) Spectrum of $H_{\rm Re}$ (in units of $\lambda$) versus $J_H/\lambda$,
with degeneracies indicated in brackets.  For $J_H\!=\!  0$, the eigenstates at
three distinct energies correspond to different ways of filling two electrons
into SO coupled single-particle states with angular momentum $j\!=\!3/2$ and
$j\!=\!1/2$.  When $\lambda\!=\!0$, we find two-particle angular momentum
eigenstates $^3P$, $^1D$, and $^1S$ in increasing order of energy. A weak SO
coupling, with $J_H/\lambda \!\gg\! 1$, splits the lowest $^3P$ manifold into
$^3P_2$, $^3P_1$, $^3P_0$. (b) Interaction dependence of total orbital and spin angular
momentum, and magnitude of SO energy. For $J_H \!\gtrsim\! \lambda$, the ground
state is composed of $L\!=\!1,S\!=\! 1$ moments which lock to yield a total angular momentum
$\vec{\cal R}=\vec L \!+\! \vec S$, with ${\cal R}\!=\!2$.}
\label{Fig:theory}
\end{figure}

On the Fe sites, a nominal valence assignment of Fe$^{3+}$ together with a
strong Hund's coupling leads to a spin $\SF \!=\!  5/2$.  On the Re sites, a
nominal valence assignment of Re$^{5+}$ ($5d^2$) leads to two electrons in the
$t_{2g}$ orbital. Thus, in contrast to the iridates, not only SOC but
also Hund's coupling ($J_H$) is important in determining the magnetic state on
Re \cite{Chen2011}. The interaction Hamiltonian together with the SOC, when projected
to the $t_{2g}$ orbital \cite{projection}, takes the form \bea H_{\rm Re} = - 2
J_H \vec S^2 - \frac{J_H}{2} \vec L^2 - \lambda (\vec \ell_1 \cdot \vec s_1 +
\vec \ell_2 \cdot \vec s_2),  \eea where $\vec S  = \vec s_1 + \vec s_2$ and
$\vec L = \vec \ell_1 + \vec \ell_2$. As seen in Fig.~\ref{Fig:theory}(a), $H_{\rm
Re}$ supports a $5$-fold degenerate ground state over a wide range of
$J_H/\lambda$ (see Supplemental Material). For $J_H/\lambda \gtrsim 1$,
Fig.\ref{Fig:theory}(b) shows that
this ground state manifold may be viewed as made up of $L \!=\!1$ and $S \!=\!
1$ moments locked into a state with total angular momentum $\vec \SR\!=\!\vec L
+ \vec S$, with $\SR \!=\! 2$.

For \BFRO, where $\lambda \!\!\gg\!\! {\cal J}$, the local moment
Hamiltonian simplifies to $H_{\rm eff} \!=\!  {\cal J}_{\rm eff} \sum_{\la
\br\br' \ra} \vec \SR_\br \cdot \vec \SF_{\br'}$, yielding an effective Heisenberg model
with moments $\SR,\SF$ on the Re and Fe sites, respectively.  We find that ${\cal
J}_{\rm eff} = {\cal J} \frac{\SR(\SR+1)+S(S+1)-L(L+1)}{2 \SR (\SR+1)}$; for
$L\!\!=\!\!S\!\!=\!\!1$, and $\SR \!=\!2$, we obtain ${\cal J}_{\rm eff} \!=\!
{\cal J}/2$. We expect that the metallic nature of \BFRO, and the concomitant
carrier delocalization, will lead to a smaller effective value of $\SF,\SR$
compared to this highly localized viewpoint.

\noindent {\bf Spin wave dispersion:}
The model $H_{\rm eff}$ has a ferrimagnetic ground state, with
antiferromagnetic order of $\vec\SF$ and $\vec \SR$ leading to a net ordered
moment. A spin wave calculation \cite{Kaplan1958} around this ground state yields two modes,
with energies $\Omega_\pm(\bq) \!=\! \sqrt{\cS_+^2 \gamma_0^2  - \SF \SR
\gamma^2_\bq} \pm \cS_-  \gamma_0$ where $\cS_\pm=(\SF\pm \SR)/2$, and
$\gamma_\bq
= 2 {\cal J}_{\rm eff} (\cos q_x + \cos q_y + \cos q_z)$, with $\gamma_0=
6{\cal J}_{\rm eff}$.
At the ordering wavevector, $\Omega_-$ has a gapless quadratic dispersion,
while $\Omega_+$ has a gap $ 2 \cS_- \gamma_0$. At $T=0$, we find the dynamic
structure factor
for transverse spin fluctuations
\bea
\!\!\!\!\!\!\!\!
S_\perp (\bq,\omega) \!&=&\! 2\pi \!\! \sum_{\sigma=\pm} \! (G_\bq \!-\! \sigma
\cS_-) \delta(\omega\!-\!
\Omega_\sigma(\bq))
\eea
where $G_\bq = (\gamma_0 \cS_+^2 - \gamma_\bq \SF \SR)/\sqrt{\cS^2_+ \gamma^2_0
- \SF\SR \gamma^2_\bq}$.  As shown in Fig.~\ref{Fig:SQW}(a-d) and
Fig.~\ref{Fig:cuts}(d), setting $\gamma_0 \SF\!=\!39$ meV and $\gamma_0
\SR\!=\!25$meV  in the theoretical plots leads to a broad agreement between the
experimental data and the powder averaged theoretical result  for
$S(\bq,\omega)$, both in the existence and dispersion of the two magnetic
modes, and in the (near) gaplessness of the lower energy mode.

\noindent {\bf Spin-orbital locking on Re:} For momenta with $\gamma_\bq\!=\! 0$, the
spin wave dispersion yields $\Omega_+/\Omega_- \!=\!  \SF/\SR$.  Since these
momenta dominate the magnon density of states, we can use the ratio of the
observed peak positions in Fig.~\ref{Fig:cuts} ($39$~meV, $25$~meV), to
deduce that $\SF/\SR \!\approx\!  1.6$. If we assume that the Re moments have a
pure spin origin, we have to set $\SR \lesssim 1$. This assumption, however,
yields a Fe spin $\SF \lesssim 1.6$, which is anomalously low --- first principles calculations \cite{Wu2001,Jeon2010},
a na\"ive valence assignment of  Fe$^{3+}$, and the measured large saturation
magnetization \cite{Prellier2000}, all point to a much larger Fe moment. Our observations thus
strongly suggest that we must have $\SR > 1$, indicating a nonzero orbital contribution
to the Re moment, in qualitative agreement with XMCD
measurements.

In order to obtain estimates of the moment sizes and the exchange
coupling, we combine our INS results with previous XMCD and magnetization
data. XMCD measurements indicate a significant static orbital contribution to the
magnetization on Re, with $\mu^{\rm orb}_{\rm Re}/\mu^{\rm spin}_{\rm Re} \!
\approx\! -0.3$.  This allows us to set $L\!\approx\! 0.6 S$, which yields $S
\!\approx\!  0.63 \SR$ and $L \!\approx\! 0.37 \SR$.  High field magnetization
measurements on \BFRO indicate a saturation magnetization $m_{\rm sat} \!
\approx\! 3 \mu_B$.  Together with our neutron data, this constrains the
moment sizes to be $\SR\!\approx\!1.3$ and $\SF \!\approx\! 2.1$, yielding an
estimated exchange coupling ${\cal J}_{\rm eff} \!\approx\! 3.1$~meV.  We have
checked that including a small direct Re-Re Heisenberg exchange $\sim 0.1 {\cal
J}_{\rm eff}$ slightly modifies the spin wave dispersion but does not
significantly affect our estimate of $\SR$.  (A large Re-Re exchange coupling
leads to a dispersion which is not consistent with our data.) Thus, while
previous XMCD measurements on \BFRO have shown that there is a static orbital
contribution to the ordered magnetic moment on Re in the ferrimagnetic state,
our work shows that such SO locked moments on Re also play a role in the low
energy magnetic {\it excitations}.

\noindent {\bf Magnetic transition temperature:} We use the above values of the
moment sizes and exchange couplings to estimate the magnetic $T_c$. The
nearest neighbor classical Heisenberg model on a three-dimensional cubic lattice,
with moments $\SF,\SR$ on the two sublattices, has a mean field transition temperature
$2 {\cal J}_{\rm eff} \SF \SR$.  Assuming a quantum renormalized
$T_c \approx 2 {\cal J}_{\rm eff} \sqrt{\SF\SR} (\sqrt{\SF\SR}+1)$, we estimate
$T_c \approx 315$K, in rough agreement with the
measured $T_c^{\rm expt} \approx 304$K. If one takes the limit of fully localized moments,
setting
$\SF=2.5$ and $\SR=2$, one obtains $T_c \approx 520$~K, remarkably
close to that of the insulating compound \CFRO\!. $T_c$ calculations retaining
the itinerant Re electrons will be reported elsewhere.

\noindent {\bf Structural transition and absence of spin gap:} \BFRO has a weak tetragonal distortion, with
$c/a\!<\! 1$, which onsets at the magnetic $T_c$ \cite{Azimonte2007}.  Since a Jahn-Teller
distortion would lead to $c/a \!>\! 1$, not necessarily coincident with $T_c$,
we ascribe this distortion to SOC.  Going beyond $H_{\rm eff}$,
we expect a term $- \epsilon \sum_\br ({\cal R}_{\br,x}^4 + {\cal R}_{\br,y}^4 +
{\cal R}_{\br,z}^4)$, arising from the cubic anisotropy, which locks the Re moment
(and thus also the Fe spins) to the crystal axes. Such a magnetostructural
locking term with $\epsilon \!>\! 0$
explains the observed tetragonal distortion at $T_c$ as arising from weak orbital order,
and would lead to a spin gap of order $\epsilon$. This locking
is expected to be small; on experimental grounds since we find no clear
evidence of a spin gap, and on theoretical grounds since it arises from a
spin-orbit induced  weak mixing of well-separated $t_{2g}$ and $e_g$ crystal field levels \cite{Chen2011}.
A small magnetostructural locking term is consistent with the measured weak coercive field
$\sim\! 0.2$ Tesla.

\noindent {\bf Summary:} We have used inelastic neutron scattering and theoretical modelling to
study the magnetic excitations in \BFRO\!\!, inferring the presence of strong
correlations and spin orbit coupled moments on Re, and obtaining a broad
understanding of the phenomenology in its ferrimagnetic state. Further
efforts are necessary to synthesize single crystals or good quality thin films of
\BFRO and other DPs.  In future work, we
will extend our experiments to other DP materials, and
incorporate the itinerant character of  Re electrons in our theoretical
modelling, both of which would lead to a better understanding of
novel 5d-based TMOs.

\begin{acknowledgements}
Work at Toronto was supported by the NSERC of Canada, the Banting Postdoctoral Fellowship program, and the Canada Research Chair program. K.W.P.
acknowledges support from the Ontario Graduate Scholarship. B.C.J. and T.W.N. are supported by
the Research Center Program of Institute for Basic Science (IBS) in Korea. Research at ORNL's Spallation Neutron Source was sponsored by the Scientific User Facilities Division, Office of Basic Energy Sciences, U.S. Department of Energy.
\end{acknowledgements}

\newpage
\section{Interaction effects: Atomic limit}

For the $d^{(2)},d^{(3)},d^{(4)}$ configuration of electrons in the $t_{2g}$
orbital, we have to consider matrix elements of the Coulomb interaction on the
same footing as the spin orbit coupling. The interaction Hamiltonian projected to
the $t_{2g}$ orbitals is given by\cite{fazekas}
\bea H_{\rm int} &=& U \sum_\alpha n^\pdg_{\alpha\upa}
n^\pdg_{\alpha\dna} + (U- 5 \frac{J_H}{2}) \sum_{\alpha<\beta} n^\pdg_\alpha
n^\pdg_\beta \nonumber \\ &-& 2 J_H \sum_{\alpha< \beta} \vec S^\pdg_\alpha
\cdot \vec S^\pdg_\beta + J_H \sum_{\alpha \neq \beta} d^\dg_{\alpha\upa}
d^\dg_{\alpha\dna} d^\pdg_{\beta\dna} d^\pdg_{\beta,\upa}.
\eea
After some algebra, this can be reexpressed in terms of
rotationally invariant operators as
\bea H_{\rm int} = \frac{U- 3 J_H}{2}
n^2_{\rm tot} - 2 J_H \vec S_{\rm tot}^2 - \frac{J_H}{2} \vec L_{\rm tot}^2
\eea
where we assume the normal ordered form of these operators. For a
$d^{(2)}$ configuration, $n_{\rm tot}=2$.  Including the spin orbit coupling
term leads to the effective atomic Hamiltonian for Re
\bea H_{\rm Re} = - 2 J_H
\vec S^2 - \frac{J_H}{2} \vec L^2 - \lambda (\vec \ell_1 \cdot \vec s_1 + \vec
\ell_2 \cdot \vec s_2)
\eea
where $\vec L = \vec \ell_1 + \vec \ell_2$ and $\vec S = \vec s_1 + \vec s_2$.
To diagonalize this Hamiltonian for a $d^{(2)}$ configuration, we write the
full Hamiltonian in the basis $|L,m_\ell,S,m_s\ra$ corresponding to total
orbital and total spin angular momentum.  Since the individual orbital angular
momenta $\ell_1 \!=\! \ell_2 \!=\! 1$ and individual spin angular momenta are
$s_1\!=\!s_2\!=\!1/2$, we use a shorthand for the Clebsch-Gordan coefficients,
defining them via
\bea |L,m_\ell,S,m_s \ra &=& |L,m_\ell\ra \otimes |S,m_s\ra
\\ |L,m_\ell\ra &=& \sum_{m_1,m_2} C^{L,m_\ell}_{m_1,m_2}  |m1,m2\ra \\
|S,m_s\ra &=& \sum_{s_1,s_2}  C^{S,m_s}_{s_1,s_2} |s1,s2 \ra
\eea
in terms of which the full Hamiltonian becomes
\bea \la L',m'_\ell,S',m'_s|
H^{(2)}_{\rm at} | L,m_\ell,S,m_s \ra  \equiv
H^{L',m'_\ell,S',m'_s}_{L,m_\ell,S,m_s}
\eea
where
\bea
&&H^{L',m'_\ell,S',m'_s}_{L,m_\ell,S,m_s} = \delta_{L,L'} \delta_{S,S'}
\delta_{m'_\ell,m_\ell} \delta_{m'_s,m_s} E_{L,S} \nonumber \\ &-&\lambda
\sum_{m_1,m_2,s_1,s_2} C^{L,m_\ell}_{m_1,m_2} C^{S,m_s}_{s_1,s_2} ( 2 m_1 s_1
\bar{C}^{L',m'_\ell}_{m_1,m_2} \bar{C}^{S',m'_s}_{s_1,s_2} \nonumber \\ &+&
\sqrt{2} \bar{C}^{L',m'_\ell}_{m_1+1,m_2} \bar{C}^{S',m'_s}_{s_1-1,s_2} +
\sqrt{2} \bar{C}^{L',m'_\ell}_{m_1-1,m_2} \bar{C}^{S',m'_s}_{s_1+1,s_2})
\eea
and
\be E_{L,S} = \left[ -2 J_H S(S+1) - \frac{J_H}{2} L(L+1) \right]
\ee
Here, we must restrict ourselves to totally antisymmetric electronic states;
$(L,S)=(0,0),(1,1),(2,0)$ yield the allowed $15$ basis states.

When $J_H\!=\! 0$, we find eigenstates at three distinct energies
corresponding to filling two electrons into single-particle states corresponding
to a low energy $j\!=\!3/2$ manifold or a higher energy $j\!=\!1/2$ doublet.
On the other
hand,  when $\lambda\!=\!0$, we find $H_{\rm Re}$ has, in increasing order of energy,
total angular momentum eigenstates $^3P$, $^1D$, and $^1S$.

The numerically computed spectrum of $H_{\rm Re}$ is shown in Fig.~4 of the
paper.  Over a wide range of $J_H/\lambda$, we find a $5$-fold degenerate
ground state when spin-orbit coupling competes with $J_H$. For $J_H/\lambda \gg
1$, we can show that the $^3P$ ground states at $\lambda=0$ split into spin-orbit
coupled states which may be labelled by total angular momentum $L+S=2,1,0$ in
increasing order of energy.
(corresponding to $^3P_2,^3P_1,^3P_0$ states with
degeneracies $5,3,1$).  This shows that a local $- \lambda \vec
L\cdot \vec S$, with $L=S=1$, is a good description of the lowest energy
manifold of states when $J_H/\lambda \gtrsim 1$.
However, when $J_H \lesssim \lambda$, this sequence
changes to $5,1,3$ (in ascending order) suggesting that such a simple
description fails.

\begin{figure*}[!] \includegraphics[]{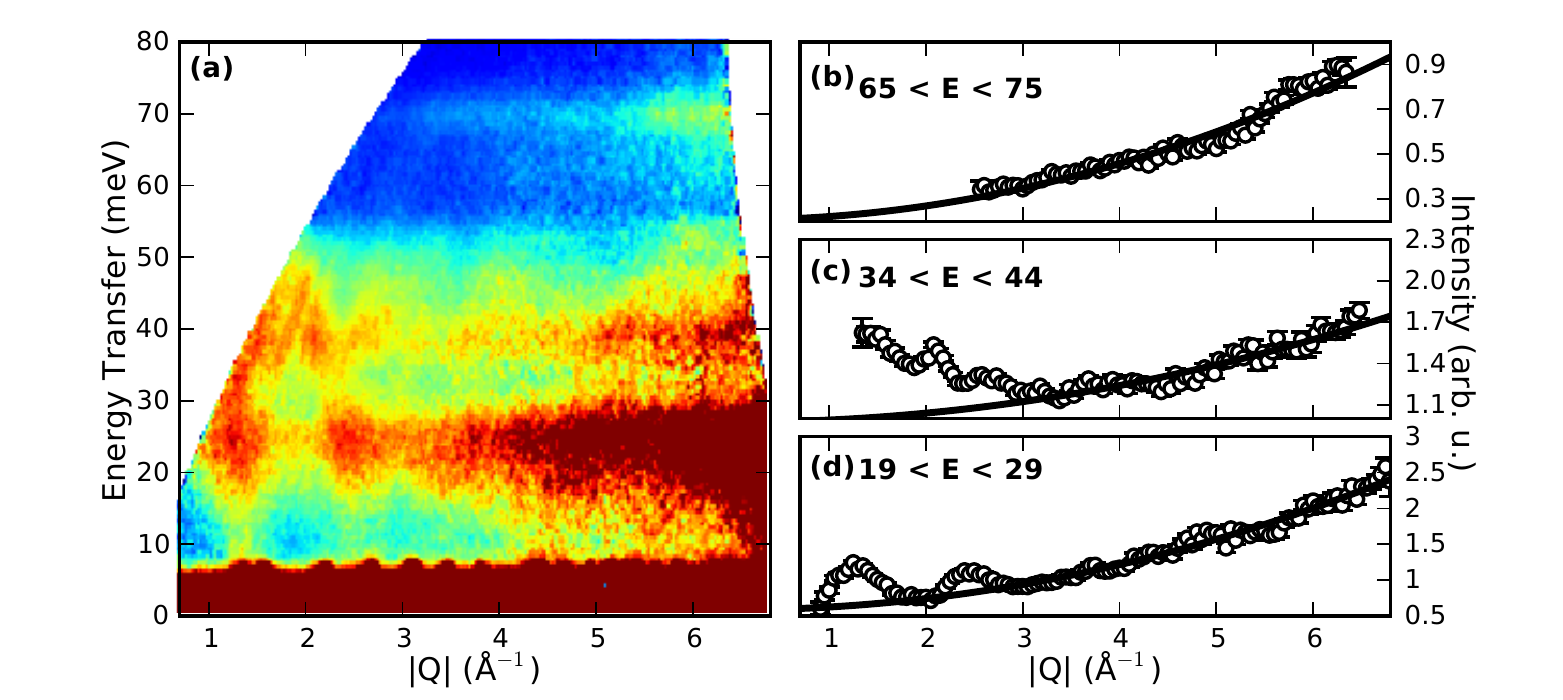}
    \caption{\label{Fig:phonon} (a) Neutron scattering intensity at 34 K for
    and incident energy of 120 meV. An empty Al can background has been
    subtracted from the data. (b)---(d) Constant energy cuts across bands of
    inelastic scattering at 70, 40 and 25 meV respectively. Solid black lines
    are a fit to $I(Q) = A(E) + BQ^2$ delimiting the Q-dependent contribution
of phonon scattering at each energy.} \end{figure*}

\section{Phonon background}
The measured scattering intensity consists of a number of components including coherent nuclear and magnetic scattering, as well as incoherent processes.  Additional background scattering  originating from the sample environment, and detector dark current is eliminated by subtracting the signal measured for an empty Al sample can using identical instrumental configuration. The signal of interest is coherent scattering from magnons, which has a momentum dependent intensity dominated by the magnetic form factor. In general, the magnetic form-factor rapidly decays as a function of Q, thus the magnetic INS intensity will decrease with increasing Q. In contrast, both coherent scattering from phonons and incoherent nuclear scattering intensities increase quadratically with Q in a powder averaged measurement \cite{Squires:78}. Any periodic modulations of the coherent phonon scattering arising from the structure factor should also increase in intensity with Q.

A map of the inelastic neutron scattering at 34 K, for 120 meV incident
energy is shown in \figref{Fig:phonon} (a). There are three bands of
inelastic scattering, around 25, 40, and 70 meV which increase in intensity
with increasing Q. We associate each of these with three phonon bands. The magnetic signal emerges from the antiferromagnetic zone center at $\mathrm{Q} = 1.35$~\AA\up{-1} and extends into two bands with maximum intensities at 25~meV and 39~meV.

To highlight the momentum dependence of the scattering intensities constant
energy cuts  through each band of inelastic scattering are shown in
\figref{Fig:phonon} (b) -- (d). Around 70 meV, [\figref{Fig:phonon} (b)], the
scattering is dominated by phonons, here the momentum dependence of
scattering intensity is entirely described by the quadratic form $I(Q) =
A(E) + BQ^2$, where $A$ is a constant function of Q parameterizing background
originating from the small multiple scattering contributions to the inelastic
scattering. On average $A$ is a decaying function of energy. In
\figref{Fig:phonon} (c) and (d) the overall intensity increases with
increasing Q at high Q, and above Q = 3~\AA\up{-1} the scattering is
dominated by phonons, as can be seen from the fits to $I(Q)$ (solid black
lines). However, below Q = 3~\AA\up{-1} the INS intensity clearly increases
above the phonon background with decreasing Q.  Furthermore in
\figref{Fig:phonon} (d) the low Q scattering intensity modulation is
consistent with the magnetic Brillouin zone. Thus, the magnetic scattering is
well separated in Q from the phonon scattering, and the magnetic scattering
is clearly identified through momentum, and temperature dependencies (see
Fig.~3 of the main text). We note that the two lower phonon
modes, which are common to many perovskite materials, are at energies which
are not far from the zone-boundary magnon mode energies. Further single
crystal inelastic neutron scattering measurements are required to determine
whether this is a mere coincidence, or a result of magnon-phonon coupling in
this material.


\begin{thebibliography}{35}
\bibitem [{pro()}]{projection}%
  \BibitemOpen
  \href@noop {} {}\bibinfo {note} {When projected to $t_{2g}$ subspace, the
  effective orbital angular momentum becomes $L_{eff}=1$ with opposite sign. We
  drop the subscript and use this $L$ in the following
  discussions.}\BibitemShut {Stop}%
\end{thebibliography}
\end{document}